\documentclass[a4paper]{jpconf}
\usepackage{graphicx}
\usepackage{amssymb,amsfonts,amsmath,,amscd}
\newcommand{\bea}{\begin{eqnarray}}
\newcommand{\eea}{\end{eqnarray}}
\newcommand{\nn}{\nonumber \\}
\begin{document}
\title{Hidden Supersymmetries of Deformed Supersymmetric Mechanics}

\author{Stepan Sidorov}

\address{Bogoliubov Laboratory of Theoretical Physics, JINR, 141980 Dubna, Moscow Region, Russia}

\ead{sidorovstepan88@gmail.com}

\begin{abstract}
We consider quantum models corresponding to superymmetrizations of the two-dimensional harmonic oscillator based on worldline $d\,{=}\,1$ realizations of the supergroup SU$(\,{\cal N}/2\,|1)$, where the number of supersymmetries ${\cal N}$ is arbitrary even number. Constructed models possess the hidden supersymmetry SU$(\,{\cal N}/2\,|2)$. Degeneracies of energy levels are spanned by representations of the hidden supersymmetry group.
\end{abstract}

\section{Introduction}

In \cite{WS, BN1, BN2} there were studied a new class of ${\cal N}\,{=}\,4$, $d\,{=}\,1$ supersymmetric quantum mechanics (SQM) models known as ``Weak Supersymmetry'' models, where the appropriate superalgebra involves, besides Hamiltonian, extra bosonic generators not commuting with supercharges. 
In \cite{DSQM, SKO} we reproduced these models, associated with the multiplets\footnote{The numbers ${\bf (k,{\cal N},{\cal N}-k)}$ display the field content of ${\cal N}$, $d\,{=}\,1$ multiplets. For example, the multiplet ${\bf (1,4,3)}$ contains ${\bf 1}$ physical bosonic field, ${\bf 4}$ physical fermionic fields and ${\bf 3}$ auxiliary bosonic fields.} ${\bf (1,4,3)}$ and ${\bf (2,4,2)}$, from a superfield approach based on the $d\,{=}\,1$ worldline supersymmetry SU(2$|$1). We identified the weak supersymmetry superalgebra with the superalgebra $su(2|1)$, where we considered it as a deformation of the standard ${\cal N}\,{=}\,4$, $d\,{=}\,1$ Poincar\'e superalgebra by a mass parameter $m$. 

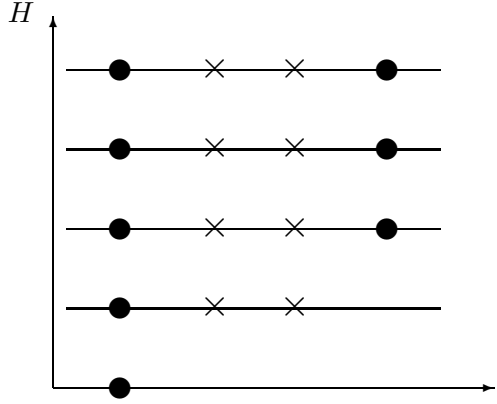
\begin{figure}[h]
\begin{center}
\begin{picture}(200,140)
\put(25,0){\line(0,1){130}}
\put(25,130){\vector(0,1){10}}

\put(8,138){$H$}

\put(25,0){\line(1,0){160}}
\put(180,0){\vector(1,0){10}}

\put(30,30){\line(1,0){140}}
\put(30,60){\line(1,0){140}}
\put(30,90){\line(1,0){140}}
\put(30,120){\line(1,0){140}}

\put(50,0){\circle*{8}}
\put(50,30){\circle*{8}}
\multiput(50,60)(100,0){2}{\circle*{8}}
\multiput(50,90)(100,0){2}{\circle*{8}}
\multiput(50,120)(100,0){2}{\circle*{8}}
\multiput(80,27)(30,0){2}{\large $\times$}
\multiput(80,57)(30,0){2}{\large $\times$}
\multiput(80,87)(30,0){2}{\large $\times$}
\multiput(80,117)(30,0){2}{\large $\times$}

\end{picture}
\end{center}
\caption{The degeneracy of energy levels of the one-dimensional SU$(2|1)$ supersymmetric harmonic oscillator drawn in \cite{WS}. Circles and crosses indicate bosonic and fermionic states, respectively.}\label{f1}
\end{figure}

The simplest model of weak supersymmetry considered in \cite{WS} corresponds to ${\cal N}\,{=}\,4$ supersymmetric extension of the one-dimensional harmonic oscillator. One of the distinct features of this model is the non-trivial degeneracy of energy levels (see Figure \ref{f1}), that was explained in \cite{DSQM} in the framework of the SU$(2|1)$ representation theory \cite{SU21reps}. The first excited level is given by unequal numbers of fermionic and bosonic states forming the fundamental
representation of SU$(2|1)$. All other higher excited levels show up the standard 4-fold degeneracy containing equal numbers of fermionic and bosonic states and corresponding to the simplest typical SU$(2|1)$ representation.

In \cite{DSQM}, we also considered the chiral multiplet ${\bf (2,4,2)}$ and constructed its general superfield action. One of the simplest models corresponds to SU$(2|1)$ supersymmetric extension of the two-dimensional harmonic oscillator and possesses the hidden supersymmetry SU$(2|2)$. Degeneracies at each energy level, except the ground state, are spanned by SU$(2|2)$ representations characterized by equal numbers of fermionic and bosonic states \cite{Beisert}.

In this paper, we consider SU$(\,{\cal N}/2\,|1)$ supersymmetric extensions of the two-dimensional harmonic oscillator, where the number of supersymmetries ${\cal N}$ is even and not restricted from above\footnote{SU$(4|1)$ superfield approach for the chiral multiplet ${\bf (2,8,6)}$  was studied in \cite{SU41}.}. We show that the presence of hidden supersymmetry is a common feature of such models.

\section{Two-dimensional harmonic oscillator}
The quantum Hamiltonian of the two-dimensional oscillator reads
\bea
	H_{\rm bos.}=\frac{1}{2}\,\sum_{a=1,2}\left(-\,\partial_a^2 + m^2 x_a^2\right)\quad\Rightarrow\quad H_{\rm bos.}=-\,\partial_z\partial_{\bar z}+m^2z\bar{z}\,,
\eea
where 
\bea
	z=\frac{1}{\sqrt{2}}\left(x_1+ix_2\right),\qquad \bar{z}=\frac{1}{\sqrt{2}}\left(x_1-ix_2\right)
\eea
The Hamiltonian can be rewritten as
\bea
	H_{\rm bos.}=a^{+}a^{-} + b^{+}b^{-}+m,
\eea
where creation and annihilation operators are given by
\bea
	&&a^+=\frac{i}{\sqrt{2}}\left(\partial_{\bar z}-mz\right),\quad
	a^-=\frac{i}{\sqrt{2}}\left(\partial_{z}+m\bar{z}\right),\qquad\left[a^-,a^+\right]=m\,,\nn
	&&b^+=\frac{i}{\sqrt{2}}\left(\partial_{z}-m\bar{z}\right),\quad
	b^-=\frac{i}{\sqrt{2}}\left(\partial_{\bar z}+mz\right),\qquad\left[b^-,b^+\right]=m\,.
\eea

Wave functions are written via the creation operator $a^{+}$ and $b^{+}$ acting on the ground state $\left|0\right\rangle=\exp{\left(-mz\bar{z}\right)}$ as
\begin{eqnarray}
	&&\left| n_1, n_2\right\rangle = \left(a^+\right)^{n_1}\left(b^+\right)^{n_2}\left|0\right\rangle,\nn
	&&\left| n_1 +1, n_2\right\rangle = a^+\left| n_1, n_2\right\rangle = \left(a^+\right)^{n_1 + 1}\left(b^+\right)^{n_2}\left|0\right\rangle,\nn
	&&\left| n_1, n_2+1\right\rangle = b^+\left| n_1, n_2\right\rangle = \left(a^+\right)^{n_1}\left(b^+\right)^{n_2+1}\left|0\right\rangle.\label{wf}
\end{eqnarray}
The operators $a^{-}$ and $b^{-}$ annihilate the ground state and reduce the number of creation operators:
\begin{eqnarray}
	&&a^-\left|0\right\rangle = 0\,,\qquad
	a^-\left| n_1, n_2\right\rangle = n_1\,m\,\left| n_1-1, n_2\right\rangle,\nn
	&&b^-\left|0\right\rangle = 0\,,\qquad
	b^-\left| n_1, n_2\right\rangle = n_2\,m\,\left| n_1, n_2-1\right\rangle.
\end{eqnarray}
The spectrum of the Hamiltonian $H_{\rm bos.}$ is then 
\begin{eqnarray}
	H_{\rm bos.}\left| n_1, n_2\right\rangle = \left(n_1 + n_2+1\right)m\,\left| n_1, n_2\right\rangle.
\end{eqnarray}

\subsection{Hidden symmetry}

One can introduce SU(2) symmetry generators commuting with the Hamiltonian:
\bea
	J_{+} = \frac{1}{m}\,b^+a^-, \qquad J_{-}=\frac{1}{m}\,a^+b^-, \qquad J_3 = \frac{1}{2m}\left(b^+b^--a^+a^-\right).\label{SU2}
\eea
They commute as
\bea
	\left[J_{+},J_{-}\right]=2J_3\,,\qquad
	\left[J_3,J_{\pm}\right]=\pm\,J_{\pm}\,.
\eea
It is none other than the well known hidden SU(2) symmetry of the two-dimensional harmonic oscillator. In general, $N$-dimensional harmonic oscillator possesses SU$(N)$ symmetry (see e.g. \cite{Perelomov}).

\begin{figure}[h]
\begin{center}
\begin{picture}(250,180)
\put(25,0){\line(0,1){170}}
\put(25,170){\vector(0,1){10}}
\put(8,0){$m$}
\put(3,40){$2m$}
\put(3,80){$3m$}
\put(3,120){$4m$}
\put(3,160){$5m$}
\put(8,178){$H$}

\put(25,0){\line(1,0){220}}
\put(240,0){\vector(1,0){10}}

\put(40,40){\line(1,0){200}}
\put(40,80){\line(1,0){200}}
\put(40,120){\line(1,0){200}}
\put(40,160){\line(1,0){200}}

\put(60,0){\circle*{8}}
\put(60,40){\circle*{8}}
\put(60,80){\circle*{8}}
\put(60,120){\circle*{8}}
\put(60,160){\circle*{8}}

\put(100,40){\circle*{8}}
\put(100,80){\circle*{8}}
\put(100,120){\circle*{8}}
\put(100,160){\circle*{8}}

\put(140,80){\circle*{8}}
\put(140,120){\circle*{8}}
\put(140,160){\circle*{8}}

\put(180,120){\circle*{8}}
\put(180,160){\circle*{8}}

\put(220,160){\circle*{8}}
\end{picture}
\end{center}
\caption{The degeneracy of energy levels of the two-dimensional oscillator. Each energy level is given by $n + 1$ states.}\label{f2}
\end{figure}
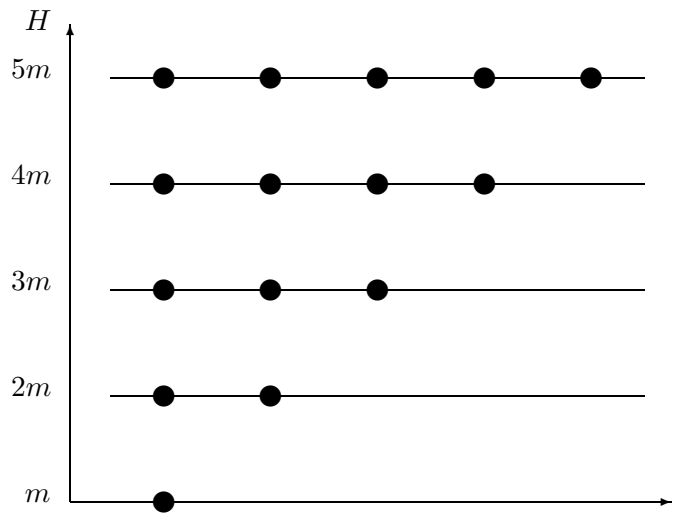

Since the generators \eqref{SU2} commute with the Hamiltonian, we have a degeneracy of energy levels corresponding to SU(2) representations.
As shown in Figure \ref{f2}, the energy level $n$ is given by $n+1$ states with the wave function defined as a sum of \eqref{wf}:
\bea
	\left|n\right\rangle = \sum^{n}_{n_1=0} C_{n_1}\left| n_1, n-n_1\right\rangle,\qquad H\,\left|n\right\rangle=\left(n+1\right)m\,\left|n\right\rangle, \qquad C_{n_1}={\rm const}.\label{wfn}
\eea
Indeed, the action of $2J_3$ on these states takes the integer eigenvalues from $-n$ to $n$.

\section{Supersymmetric two-dimensional harmonic oscillator}
Supersymmetric Weyl-ordered Hamiltonian reads
\bea    
	H=a^+a^-+b^+b^-+m\,\psi^K\bar{\psi}_K + m\left(1-\frac{{\cal N}}{4}\right),\label{H}
\eea 
where the fermionic fields $\psi^I$ and $\bar{\psi}_J$ anticommute as
\begin{eqnarray}
	\left\lbrace\psi^I,\bar{\psi}_J\right\rbrace = \delta^I_J\,.
\end{eqnarray}
The capital indices $I,J,K,L$ refer to the SU$(\,{\cal N}/2\,)$ fundamental and anti-fundamental representations. The supersymmetric Hamiltonian \eqref{H} commute with supercharges defined as
\bea
    Q^I = \sqrt{2}\,\psi^I\,a^-,\qquad
    \bar{Q}_J = \sqrt{2}\,\bar{\psi}_J\,a^+.\label{Q}
\eea
They close on the centrally-extended superalgebra $\widehat{su}(\,{\cal N}/2\,|1)$ given by the following non-vanishing (anti)commutators:
\bea
    &&\left\lbrace Q^{I}, \bar{Q}_{J}\right\rbrace = 2m\,L^{I}_{J}-2m\,\delta^{I}_{J}R + \delta^{I}_{J}H,\qquad
    \left[L^I_J,  L^K_L\right]
    = \delta^K_J L^I_L - \delta^I_L L^K_J,\nn
    &&\left[L^I_J, Q^{K}\right]
    = \delta^K_J Q^{I} - \frac{2}{{\cal N}}\,\delta^I_J Q^{K} ,\qquad
    \left[L^I_J, \bar{Q}_{L}\right] = \frac{2}{{\cal N}}\,\delta^I_J\bar{Q}_{L}-\delta^I_L\bar{Q}_{J}\,,\nn
    &&\left[R, Q^{I}\right]=\left(1-\frac{2}{{\cal N}}\right)Q^{I},\qquad \left[R, \bar{Q}_{J}\right]=-\left(1-\frac{2}{{\cal N}}\right)\bar{Q}_{J}\,.
\eea
The ${\rm SU(\,{\cal N}/2\,)}\times{\rm U}(1)$ subgroup generators $L^I_J$ and $R$ are written as
\bea      
    &&L^I_J =\psi^I\bar{\psi}_J - \frac{\delta^I_J}{4}\,\psi^K\bar{\psi}_K\,,\nn
    &&R = \frac{1}{2m}\left(b^+b^--a^+a^-\right)+\left(\frac{1}{2}-\frac{2}{{\cal N}}\right)\psi^K\bar{\psi}_K-\frac{{\cal N}}{4}\left(\frac{1}{2}-\frac{2}{{\cal N}}\right).
\eea
The mass-dimensional generator $H$, associated with Hamiltonian, is a central charge generator. In the limit $m=0$, the ${\rm SU}(\,{\cal N}/2\,)\times {\rm U}(1)$ generators become automorphism generators of the standard ${\cal N}$, $d\,{=}\,1$ Poincar\'e superalgebra.

\subsection{Hidden supersymmetry}
One can check the hidden SU(2) symmetry generators \eqref{SU2} commutes with the supersymmetric Hamiltonian. However their commutators with \eqref{Q} give new supercharges
\bea
	\left[J_-, Q^I\right] =-\,S^I,\qquad \left[J_+, \bar{Q}_{J}\right] =\bar{S}_J\,.
\eea
The new supercharges are written as
\bea
	S^I =\sqrt{2}\,\psi^I\,b^{-},\qquad 
	\bar{S}_J=\sqrt{2}\,\bar{\psi}_J\,b^{+}.
\eea
Thus, the superalgebra \eqref{su21} is extended by the new supercharges and SU(2) generators.

Splitting the U(1) generator $R$ as
\bea
	R = J_3 +\left(1-\frac{4}{{\cal N}}\right)F\quad\Rightarrow\quad F = \frac{1}{2}\,\psi^K\bar{\psi}_K-\frac{{\cal N}}{8}\,,\qquad J_{3} = \frac{1}{2m}\left(b^+b^--a^+a^-\right),
\eea
we obtain the centrally-extended superalgebra $\widehat{su}(\,{\cal N}/2\,|2)$ given by
\bea
	&&\left\lbrace Q^{I}, \bar{Q}_{J}\right\rbrace = 2m\,L^{I}_{J}-2m\,\delta^{I}_{J}J_3 + 2m\left(\frac{4}{{\cal N}}-1\right)\delta^{I}_{J}F + \delta^{I}_{J}H,\qquad\lbrace Q^I,\bar{S}_J\rbrace =2m\,\delta_J^I J_+\,,\nn
    &&\left\lbrace S^{I}, \bar{S}_{J}\right\rbrace = 2m\,L^{I}_{J}+2m\,\delta^{I}_{J}J_3+2m\left(\frac{4}{{\cal N}}-1\right)\delta^{I}_{J}F + \delta^{I}_{J}H,\qquad\lbrace S^I,\bar{Q}_J\rbrace =2m\,\delta_J^I J_-\,,\nn
    &&\left[L^i_j, \bar{Q}_{K}\right] =\frac{2}{{\cal N}}\,\delta_J^I\bar{Q}_{K} -\delta^I_K\bar{Q}_{J}\,,\qquad
    \left[L_J^I,Q^{K}\right] = \delta^K_J Q^{I} - \frac{2}{{\cal N}}\,\delta_J^I Q^{K},\nn
    &&\left[L_J^I \bar{S}_{K}\right] =\frac{2}{{\cal N}}\,\delta_J^I \bar{S}_{K} -\delta^I_K\bar{S}_{J}\, ,\qquad
    \left[L_J^I,S^{k}\right] = \delta^K_J S^{I} - \frac{2}{{\cal N}}\,\delta^I_J S^{K},\nn
    &&\left[F, Q^{I}\right]=\frac{1}{2}\,Q^{I},\quad\left[F, \bar{Q}_{J}\right] = -\,\frac{1}{2}\,\bar{Q}_{J}\,,\quad \left[F, S^{I}\right] =\frac{1}{2}\,S^{I},\quad \left[F, \bar{S}_{J}\right]= -\,\frac{1}{2}\,\bar{S}_{J}\,,\nn
    &&\left[J_3, Q^{I}\right]=\frac{1}{2}\,Q^{I},
    \quad \left[J_3, \bar{Q}_{J}\right] =-\,\frac{1}{2}\,\bar{Q}_{J}\,,\quad\left[J_+, \bar{Q}_{J}\right] =\bar{S}_J\,,\quad\left[J_+, S^I\right] =-\,Q^I,\nn
    &&\left[J_3, S^{I}\right]=-\,\frac{1}{2}\,S^{I},
    \quad \left[J_3, \bar{S}_{J}\right] =\frac{1}{2}\,\bar{S}_{J}\,,\quad\left[J_-, \bar{S}_{J}\right] =\bar{Q}_J\,,\quad\left[J_-, Q^I\right] =-\,S^I,\nn
    &&\left[L_J^I,L^K_L\right] =  \delta^K_J L^I_L- \delta^I_L L^K_J \,,\qquad \left[J_+,J_-\right]
    =2J_3\,,\qquad \left[J_3,J_\pm\right]=\pm\,J_\pm\,.
\eea

\subsection{{\cal N}=4 supersymmetric extension}
As instructive example, let us consider in details ${\cal N}\,{=}\,4$ supersymmetric extension. Our studies of SU(2$|$1) supersymmetric mechanics \cite{DSQM} were based on the deformation of the standard ${\cal N}\,{=}\,4$, $d\,{=}\,1$ Poincar\'e superalgebra to the centrally-extended superalgebra $\widehat{su}(2|1)$ given by\footnote{The superalgebra relations here differ from those in \cite{DSQM} such that $H\rightarrow 2H$.}
\bea
    &&\lbrace Q^{i}, \bar{Q}_{j}\rbrace = 2m\,I^i_j -2m\,\delta^i_j R + \delta^i_j H,\qquad\left[I^i_j,  I^k_l\right]
    = \delta^k_j I^i_l - \delta^i_l I^k_j\,,\nn
    &&\left[I^i_j, \bar{Q}_{l}\right] = \frac{1}{2}\,\delta^i_j\bar{Q}_{l}-\delta^i_l\bar{Q}_{j}\, ,\qquad \left[I^i_j, Q^{k}\right]
    = \delta^k_j Q^{i} - \frac{1}{2}\,\delta^i_j Q^{k},\nn
    &&\left[R, Q^{k}\right]=\frac{1}{2}\,Q^{k},\qquad   
    \left[R, \bar{Q}_{l}\right]=-\,\frac{1}{2}\,\bar{Q}_{l}\,.\label{su21}
\eea
The indices $i$, $j$ ($i=1, 2$) are SU(2) indices. The generators $I^i_j$ and $R$ correspond to the ${\rm SU}(2)\times {\rm U}(1)$ subgroup. 

The quantum SU(2$|$1) generators are written as
\bea
	&&Q^i =\sqrt{2}\,\psi^i\,a^{-},\qquad 
	\bar{Q}_j=\sqrt{2}\,\bar{\psi}_j\,a^{+},\nn
	&&H=a^+a^-+b^+b^-+m\,\psi^k\bar{\psi}_k\,,\nn
	&&R = \frac{1}{2m}\left(b^+b^- - a^+a^-\right)\quad \Rightarrow \quad R \equiv J_3\,,\nn
	&&I^i_j =\psi^i\bar{\psi}_j - \frac{\delta^i_j}{2}\,\psi^k\bar{\psi}_k\,.
\eea
It is straightforward to check that they satisfy the superalgebra relation \eqref{su21}.

As was shown in \cite{DSQM} the model possesses hidden supersymmetry SU$(2|2)$.
The corresponding centrally-extended superalgebra $\widehat{su}(2|2)$ is given by
\bea
    &&\lbrace Q^i,\bar{Q}_j\rbrace =\delta_j^i H + 2m\left(  I_j^i - \delta_j^i J_3\right),
    \qquad\lbrace Q^i,\bar{S}_j\rbrace =2m\,\delta_j^i J_+\,,\nn
    &&\lbrace S^i,\bar{S}_j\rbrace = \delta_j^i H + 2m\left(  I_j^i + \delta_j^i J_3\right),
    \qquad\lbrace S^i,\bar{Q}_j\rbrace =2m\,\delta_j^i J_-\,,\nn
    &&\left[I^i_j, \bar{Q}_{l}\right] =\frac{1}{2}\,\delta^i_j\bar{Q}_{l} -\delta^i_l\bar{Q}_{j}\,,\qquad
    \left[I^i_j,Q^{k}\right] = \delta^k_j Q^{i} - \frac{1}{2}\,\delta^i_j Q^{k},\nn
    &&\left[I^i_j, \bar{S}_{l}\right] =\frac{1}{2}\,\delta^i_j\bar{S}_{l} -\delta^i_l\bar{S}_{j}\,,\qquad
    \left[I^i_j,S^{k}\right] = \delta^k_j S^{i} - \frac{1}{2}\,\delta^i_j S^{k},\nn
    &&\left[J_3, Q^{i}\right]=\frac{1}{2}\,Q^{i},
    \quad \left[J_3, \bar{Q}_{j}\right] =-\,\frac{1}{2}\,\bar{Q}_{j}\,,\quad\left[J_+, \bar{Q}_{j}\right] =\bar{S}_j\,,\quad\left[J_+, S^i\right] =-\,Q^i,\nn
    &&\left[J_3, S^{i}\right]=-\,\frac{1}{2}\,S^{i},
    \quad \left[J_3, \bar{S}_{j}\right] =\frac{1}{2}\,\bar{S}_{j}\,,\quad\left[J_-, \bar{S}_{j}\right] =\bar{Q}_j\,,\quad\left[J_-, Q^i\right] =-\,S^i,\nn
    &&\left[I^i_j,I^k_l\right] =  \delta^k_j I^i_l- \delta^i_l I^k_j \,,\qquad \left[J_+,J_-\right]
    =2J_3\,,\qquad \left[J_3,J_\pm\right]=\pm\,J_\pm\,.\label{su22}
\eea

\subsubsection{Dual superfield description.} 

Corresponding Lagrangian of the two-dimensional SU$(2|1)$ supersymmetric harmonic oscillator reads
\bea
    {\cal L}_{\rm \;on-shell} = \dot{z}\dot{\bar{z}} + \frac{i}{2}\left(\psi^i\dot{\bar{\psi}}_i-\dot{\psi^i}\bar{\psi}_i\right) - m^2z\bar{z} - m\,\psi^k\bar{\psi}_k\,.
\eea
This on-shell Lagrangian can be constructed from superfield approach in two ways, which is exceptional for ${\cal N}\,{=}\,4$ supersymmetric extension.

First way is given in the framework of SU$(2|1)$ superfield approach elaborated in \cite{DSQM}. Simple superfield Lagrangian for the chiral multiplet ${\bf (2, 4, 2)}$ reads
\bea
    {\cal L}_{\rm \;deformed}=\int d^2\vartheta\, d^2\bar{\vartheta}\left(1 + 2m\,\bar\vartheta^k \vartheta_k\right)\Psi\,\bar{\Psi}.
\eea
The second superfield construction is based on the chiral multiplet of non-deformed (standard) SQM, where the parameter $m$ introduced via superpotential term:
\bea
    {\cal L}_{\rm \;non-deformed}= \int d^2\theta\, d^2\bar{\theta}\,\Phi\,\bar{\Phi}
    +\frac{m}{2}\left[\int d^2\theta \,\Phi^{2}+ \int d^2\bar{\theta} \,\bar{\Phi}^{2}\right].
\eea
After elimination of auxiliary fields in both actions, we set equivalence of these Lagrangians. The superalgebra \eqref{su21} and the standard ${\cal N}\,{=}\,4$, $d\,{=}\,1$ superalgebra are realized on the same on-shell set $(z, \bar z, \psi^i, \bar\psi_i)$. The superalgebra \eqref{su22} is recovered as the closure of these two superalgebras. 

Define the new supercharges
\begin{eqnarray}
	\Pi^i =\frac{1}{\sqrt{2}}\left(Q^i+\bar{S}^i\right),\qquad 
	\bar{\Pi}_j=\frac{1}{\sqrt{2}}\left(\bar{Q}_j - S_j\right).
\end{eqnarray}
One can check that they form the standard ${\cal N}\,{=}\,4$ superalgebra,
\begin{eqnarray}
	\left\lbrace \Pi^i, \bar{\Pi}_j\right\rbrace = \delta^i_j H.
\end{eqnarray}
Thus, it is a subalgebra of \eqref{su22}, while the generators \eqref{SU2} become external automorphism generators:
\bea
    &&\left[J_3, \bar{\Pi}_{j}\right] =-\,\frac{1}{2}\,\bar{\Pi}_{j}\,,\qquad \left[J_3, \Pi^{i}\right]=\frac{1}{2}\,\Pi^{i},\nn
    &&\left[J_+, \bar{\Pi}_{j}\right] =\Pi_j\,,\qquad \left[J_-, \Pi^i\right] =\bar{\Pi}^i.
\eea

\subsubsection{Wave functions.} We construct supersymmetric wave functions in terms of \eqref{wfn} satisfying the same spectrum
\begin{eqnarray}
	H\left|n\right\rangle = n\,m\,\left|n\right\rangle.
\end{eqnarray} 
We impose the standard physical condition on the bosonic states $|n\rangle$ as
\bea
    \bar{\psi}_j\,|n\rangle =0\,.
\eea
Then, the fermionic states $\psi_k\,\left|n\right\rangle$ have shifted spectrum $\left(n + 1\right)m$\,:
\begin{eqnarray}
	H\,\psi_k\,\left|n\right\rangle = \left(n + 1\right)m\,\psi_k\,\left|n\right\rangle.
\end{eqnarray}
Additional bosonic states defined as $\psi_k\psi^k\,\left|n\right\rangle$ also have shifted spectrum $\left(n + 2\right)m$\,:
\begin{eqnarray}
	H\,\psi_k\psi^k\,\left|n\right\rangle = \left(n + 2\right)m\,\psi_k\psi^k\,\left|n\right\rangle.
\end{eqnarray}

\begin{figure}[h]
\begin{center}
\begin{picture}(420,180)
\put(25,0){\line(0,1){170}}
\put(25,170){\vector(0,1){10}}
\put(11,0){$0$}
\put(8,40){$m$}
\put(3,80){$2m$}
\put(3,120){$3m$}
\put(3,160){$4m$}
\put(8,178){$H$}
\put(25,0){\line(1,0){390}}
\put(410,0){\vector(1,0){10}}

\put(30,40){\line(1,0){380}}
\put(30,80){\line(1,0){380}}
\put(30,120){\line(1,0){380}}
\put(30,160){\line(1,0){380}}

\put(40,0){\circle*{7}}
\put(40,40){\circle*{7}}
\multiput(40,80)(60,0){2}{\circle*{7}}
\multiput(40,120)(60,0){2}{\circle*{7}}
\multiput(40,160)(60,0){2}{\circle*{7}}
\multiput(55,37)(20,0){2}{\large $\times$}
\multiput(55,77)(20,0){2}{\large $\times$}
\multiput(55,117)(20,0){2}{\large $\times$}
\multiput(55,157)(20,0){2}{\large $\times$}

\multiput(115,33)(0,10){14}{\line(0,1){5}}

\put(130,40){\circle*{7}}
\put(130,80){\circle*{7}}
\multiput(130,120)(60,0){2}{\circle*{7}}
\multiput(130,160)(60,0){2}{\circle*{7}}
\multiput(145,77)(20,0){2}{\large $\times$}
\multiput(145,117)(20,0){2}{\large $\times$}
\multiput(145,157)(20,0){2}{\large $\times$}

\multiput(205,73)(0,10){10}{\line(0,1){5}}

\put(220,80){\circle*{7}}
\put(220,120){\circle*{7}}
\multiput(220,160)(60,0){2}{\circle*{7}}
\multiput(235,117)(20,0){2}{\large $\times$}
\multiput(235,157)(20,0){2}{\large $\times$}

\multiput(295,113)(0,10){6}{\line(0,1){5}}

\put(310,120){\circle*{7}}
\put(310,160){\circle*{7}}
\multiput(325,157)(20,0){2}{\large $\times$}

\multiput(385,153)(0,10){2}{\line(0,1){5}}

\put(400,160){\circle*{7}}

\end{picture}
\end{center}
\caption{The degeneracy of energy levels of the two-dimensional SU$(2|1)$ supersymmetric oscillator. There are $4n$ states at the excited energy level $n$. This picture can be interpreted as a multiplication of the pictures drawn in Figures \ref{f1} and \ref{f2}. One can easily see that each tower divided by dashed line is a tower of the same type drawn in Figure \ref{f1}.}\label{f3}
\end{figure}
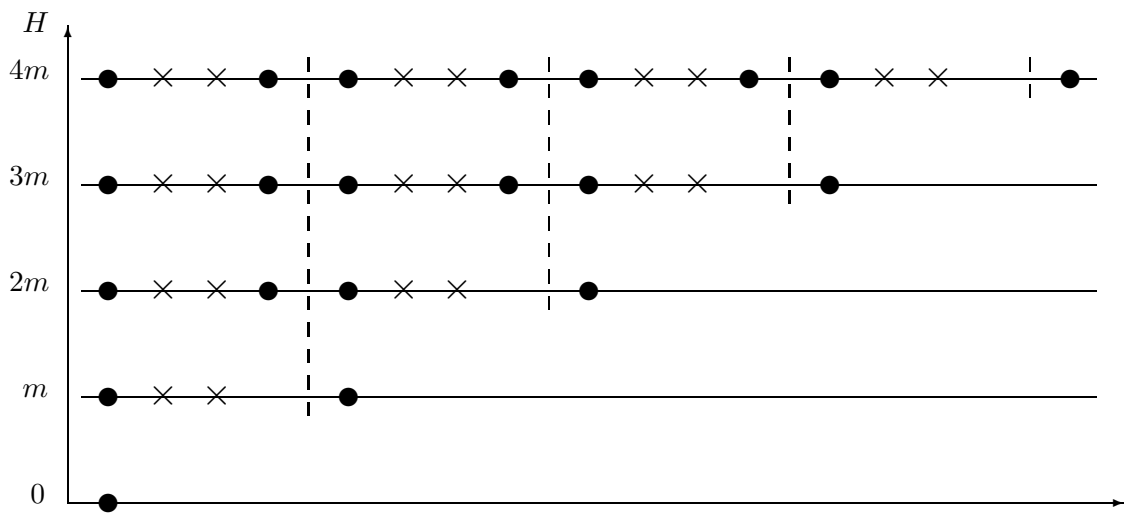

As shown in Figure \ref{f3}, each excited energy level is given by equal numbers of fermionic and bosonic states. States on the excited level $n$ form a short SU$(2|2)$ representation of the dimension $4n$ \cite{Beisert}. Note that the first excited level corresponds to the fundamental representation of SU$(2|2)$.

\section{Conclusions}
We considered superextensions of the two-dimensional harmonic oscillator with the relevant supersymmetry group SU$(\,{\cal N}/2\,|1)$. We performed quantization and constructed the Hilbert space of wave functions in terms of wave functions of bosonic harmonic oscillator. We showed that the hidden on-shell symmetry SU$(2)$ of the two-dimensional harmonic oscillator reveals the hidden supersymmetry SU$(\,{\cal N}/2\,|2)$ defining degeneracies of energy levels.
The case of ${\cal N}\,{=}\,4$ superextension is exceptional, since it has dual superfield description.

Another feature of the two-dimensional harmonic oscillator is the presence of conformal SO$(2,1)$ invariance of the trigonometric type \cite{PP}, \cite{HT}, where superconformal Hamiltonian must be an even function of the deformation parameter $m$ \cite{ISTconf}. Thus, the Hamiltonian is redefined as
\bea
	H = {\cal H}_{\rm conf.}+2mF,\qquad {\cal H}_{\rm conf.}=-\,\partial_z\partial_{\bar z}+m^2z\bar{z}=a^{+}a^{-} + b^{+}b^{-}+m.
\eea
Then the closure of the relevant superconformal group SU$(\,{\cal N}/2\,|1,1)$ and the hidden symmetry SU$(2)$ yields some extended conformal supersymmetry corresponding to the superalgebra $\sim osp({\cal N}|4)$, where the extra supercharges are written as
\bea
	&&{\cal Q}^I =\sqrt{2}\,\psi^I\,a^{+},\qquad 
	\bar{\cal Q}_J=\sqrt{2}\,\bar{\psi}_J\,a^{-},\nn
	&&{\cal S}^I =\sqrt{2}\,\psi^I\,b^{+},\qquad 
	\bar{{\cal S}}_J=\sqrt{2}\,\bar{\psi}_J\,b^{-}.
\eea
It can be interpreted as a spectrum-generating supersymmetry.

It would be interesting to study superextensions of three- or four-dimensional harmonic oscillators for hidden supersymmetries. For example, the multiplet $({\bf 4, 4, 0})$ can describe ${\cal N}\,{=}\,4$ superextension of the four-dimensional harmonic oscillator \cite{DHSS}.

\ack
The author thanks the organizers of the conference ISQS26 for the kind hospitality in Prague. He also thanks Evgeny Ivanov and Gor Sarkissian for discussions. This paper was supported by the RFBR Grant No. 18-02-01046, and by the Ministry of Science and Higher Education of Russian Federation, project No. 3.1386.2017.

\end{document}